\documentclass[twocolumn]{jpsj2} 
\usepackage{here}
\def\cepdal{CePd$_5$Al$_2$}
\def\ypdal{YPd$_5$Al$_2$}
\def\mb{${\mu}_{\rm B}$}
\def\mp{${\mu}_{\rm eff}$}
\def\tno{$T_{\rm N1}$}
\def\tnt{$T_{\rm N2}$}

\def\tt{$T_{\rm t}$}

\title{Giant Uniaxial Anisotropy in the Magnetic and Transport Properties of {\cepdal}}

\author{Takahiro \textsc{Onimaru}$^{1}$\thanks{E-mail address: onimaru@hiroshima-u.ac.jp}, Yukihiro \textsc{F. Inoue}$^{1}$, Keisuke \textsc{Shigetoh}$^{1}$, Kazunori \textsc{Umeo}$^{2}$, \\
Hirokazu \textsc{Kubo}$^{1}$, Raquel \textsc{A. Ribeiro}$^{3}$, Akihiro \textsc{Ishida}$^{1}$, Marcos \textsc{A. Avila}$^{1}$, \\
Kenji \textsc{Ohoyama}$^{4}$, Masafumi \textsc{Sera}$^{1}$ and Toshiro \textsc{Takabatake}$^{1,3}$}

\inst{$^{1}$Department of Quantum Matter, Graduate School of Advanced Sciences of Matter, Hiroshima University, Higashi-Hiroshima, Hiroshima 739-8530 \\
$^{2}$Cryogenics and Instrumental analysis Division, N-BARD, Hiroshima University, Higashi-Hiroshima, Hiroshima 739-8526 \\
$^{3}$Institute for Advanced Materials Research, Hiroshima University, Higashi-Hiroshima, Hiroshima 739-8530\\
$^{4}$Institute for Materials Research, Tohoku University, Sendai 980-8577
}

\abst{Electrical resistivity ${\rho}$, magnetic susceptibility ${\chi}$, magnetization $M$ and specific heat measurements are reported on a singlecrystalline sample of {\cepdal}, showing successive antiferromagnetic orderings at {\tno}=4.1 K and {\tnt}$=$2.9 K. The temperature dependence of ${\rho}$ shows a Kondo metal behavior with large anisotropy, ${\rho}_c$$/$${\rho}_a$ $=$3.2 at 20 K, and opening of a superzone gap along the tetragonal \textbf{\textit c}-direction below {\tno}. Both {\tno} and {\tnt} gradually increase with applying pressure up to 2.5 GPa. The data of ${\chi}(T)$ and $M$($B$) in the paramagnetic state were analyzed using a crystalline electric field (CEF) model. It led to a Kramers doublet ground state with wave functions consisting primarily of  $\left | {\pm}\frac{5}{2} \right \rangle$, whose energy level is isolated from the excited states by 230 and 300 K. This CEF effect gives rise to the large anisotropy in the paramagnetic state. In the ordered state, the uniaxial magnetic anisotropy is manifested as $M_c$$/$$M_a$$=$20 in $B$$=$5 T and at 1.9 K, and ${\chi}_{c}$/${\chi}_{a}$$=$25 in $B$$=$0.1 T and at 4 K. This huge uniaxial magnetic anisotropy in the antiferromagnetic states can be interpreted in terms of isotropic magnetic interaction among the Ce$^{3+}$ moments governed by the strong CEF. In powder neutron diffraction experiments, magnetic reflections were observed owing to the antiferromagnetic ordered states below {\tno}, however, no additional reflection was found below {\tnt}. }

\kword{{\cepdal}, Uniaxial anisotropy, Crystal electric field, Superzone gap,  Kondo lattice, Neutron diffraction}

\begin{document}
\maketitle

\section{Introduction} 

Recently there has been considerable interest in the rare-earth- and actinide-based ternary intermetallic compounds, since they have been found to have various types of ground states such as magnetic long-range order, Kondo lattice, heavy fermion state, multipole order, superconductivity, etc. In particular, the ternary R-Pd-Al (R: rare-earth or actinide) system has attracted much attention, because it includes many exotic compounds exhibiting attractive properties. For instance, UPd$_2$Al$_3$ crystallizing in the hexagonal PrNi$_2$Al$_3${-}type structure exhibits superconductivity at $T_{\rm C}$$=$2 K and antiferromagnetic (AF) ordering at $T_{\rm N}$$=$14 K, both coexisting in the ground state.\cite{Geibel90} CePd$_2$Al$_3$, isostructural to UPd$_2$Al$_3$, does not show superconductivity but does show AF transition at $T_{\rm N}$$=$2.8 K.\cite{Kitazawa92}  The magnetic ground state can be easily controlled by applying pressure\cite{Tang96} or substituting the Pd site with other elements such as Ni\cite{Sereni06}, Cu\cite{Sun03} and Ag\cite{Sun06}. Thereby, the hybridization of the 4$f$ electron state near the Fermi level with $d$ bands is changed. Moreover, the ground state of CePd$_2$Al$_3$ depends on quality of samples, for example, the AF ordering emerges in a polycrystal, whereas no phase transition is found in a single crystal.\cite{Mentink94a,Mentink94b,Tou94,Nemoto96} Another example is CePdAl which crystallizes in the ZrNiAl-type hexagonal structure, where Ce ions form a quasi-Kagome lattice on the basal plane. CePdAl undergoes an AF ordering at $T_{\rm N}$$=$2.7 K,\cite{Hulliger93,Schank94,Kitazawa94} whose magnetic structure is strongly influenced by the magnetic frustration effect.\cite{Donni96}

	Very recently, the first neptunium-based superconductor NpPd$_5$Al$_2$ with $T_{\rm C}{=}$4.9 K was discovered by Aoki {\it et al}.\cite{Aoki07} This compound crystallizes in the tetragonal ZrNi$_2$Al$_5$-type structure of the space group $I$4/$mmm$. Subsequently, Ribeiro {\it et al}. have reported the synthesis and physical properties of an isostructual cerium-based compound {\cepdal}\cite{Ribeiro07}, which is a new compound in the ternary R-Pd-Al family. The polycrystalline sample already shows a high degree of \textit{\textbf c}-axis preferred orientation. Accordingly, the electrical resistivity ${\rho}$($T$) exhibits an anisotropic behavior as ${\rho}_{\perp}$$=$1.8${\rho}_{||}$ at 300 K, where ${\rho}_{\perp}$ and ${\rho}_{||}$ are those for the current directions perpendicular and parallel to the grain length, respectively. At temperatures ranging from 10 K to 4 K, it shows ${-}$ln$T$ dependence as is seen in Kondo lattice compounds. Furthermore, ${\rho}(T)$ drops at {\tno}$=$3.9 K and bends at {\tnt}$=$2.9 K. Because two transitions were also observed in the magnetic susceptibility ${\chi}(T)$ and specific heat $C$($T$), they were ascribed to AF transitions. The effective magnetic moment of {\mp}$=$2.53$\sim$2.56 {\mb}/f.u. indicated that the Ce ion should be trivalent. In the field dependence of the magnetization $M$($B$) at 1.8 K, a metamagnetic behavior appears at  $B$$=$0.9 T only in the magnetic fields perpendicular to the grain length. The total entropy estimated from the temperature dependence of the specific heat reaches 80\% of $R$ln2 at {\tno}, suggesting that a Kramers doublet in the CEF ground state should be responsible to the magnetic transitions. From a linear extrapolation of $C$/$T$ versus $T^2$ plot between 20 K and 8 K, the electronic specific heat coefficient $\gamma$ was estimated to be 60 mJ/mol K$^2$, indicating that {\cepdal} may be classified as a strongly correlated electron system with moderate mass enhancement. 
	
	These macroscopic measurements on the polycrystalline sample suggested the strong magnetic and transport anisotropy in {\cepdal}. However, they do not allow detailed conclusion on the relation between the anisotropy and the crystal directions. In aiming at establishing the origin of the anisotropy, we have grown singlecrystalline {\cepdal} and measured ${\rho}$($T$), ${\chi}$($T$) and $M$($B$) along the \textbf{\textit a}- and \textbf{\textit c}-axes. In Ce-based compounds as well as other rare-earth compounds, the CEF effects are important for the magnetic anisotropy. We analyzed the magnetic data using the CEF model, which allowed us to determine the CEF level scheme and the 4$f$ wave functions. In order to determine the magnetic structure, neutron diffraction experiments were done on a powdered sample.

\section{Experimental Details}
\subsection{Single crystal growth and characterization}

	A singlecrystalline sample of {\cepdal} was grown by the Czochralski method using a high frequency furnace.\cite{Nakamoto95} Staring elements were 4N Ce prepared by the Ames Laboratory, 4N Pd and 5N Al. A mixture of these elements was melt and homogenized in a water-cooled Hukin-type crucible. A tungsten rod was used as a virtual seed, which was immersed to the molten sample in a hot tungsten crucible. The rod was pulled at a speed of about 8 mm/h in a purified Ar atmosphere. The obtained crystal of 60 mm in length and 5 mm in diameter was characterized by electron-probe microanalysis (EPMA), where no secondary phase was found and compositions were 1.00(1):5.00(1):1.98(1). The lattice parameters at room temperature were refined to be $a$$=$4.1666(3) ${\rm \AA}$ and $c$$=$14.942(2) ${\rm \AA}$ by the Rietveld analysis of the powder X-ray diffraction patterns using the program RIETAN-2000.\cite{Izumi00} The orientation of the singlecrystalline sample was determined by the back Laue method using an imaging plate camera, IPXC/B (TRY-SE).
	
	Polycrystalline samples of {\cepdal} and {\ypdal} were prepared by a conventional argon arc technique. For ensuring homogeneity, each ingot was turned over and remelted several times. The samples were sealed in an evacuated quartz tube and annealed at 800 $^{\circ}$C for 7 days. The $a$ and $c$ parameters of {\ypdal} were 4.1214(3) ${\rm \AA}$ and 14.822(1) ${\rm \AA}$, respectively.

\subsection{Measurements}

	Electrical resistance was measured under ambient pressure as well as applied pressures up to 2.5 GPa by a standard four-probe AC method in home-built systems. A $^3$He refrigerator and Gifford-McMahon type refrigerator were used, respectively, in the temperature ranges 0.3-200 K and 3-380 K. The pressure was generated by a clamp-type piston-cylinder pressure cell by using Daphne oil as pressure transmitting medium. The pressure was determined by the change in superconducting transition temperature of Sn. Specific heat was measured by a relaxation method using a commercial calorimeter (Quantum Design PPMS) at temperatures between 0.4 K and 300 K. Magnetization was measured using a commercial SQUID magnetometer (Quantum Design MPMS) for 1.9-350 K and in magnetic fields up to 5 T. A sample-extraction magnetometer was also used to extend the magnetic field range up to 14.5 T.  
To perform magnetization measurements at low temperatures down to 0.35 K,  we adopted a capacitive Faraday method using a high-resolution capacitive force-sensing device installed in a $^{3}$He refrigerator.\cite{Sakaki00}
Magnetic fields up to 9.5 T and field gradient of 10 T/m were applied independently by a superconducting magnet with built-in gradient coils.
	
	Powder neutron diffraction experiments were performed using the high efficiency and resolution powder diffractometer, HERMES, of Institute for Materials Research (IMR), Tohoku University, installed at the JRR-3M reactor in Japan Atomic Energy Agency (JAEA), Tokai.\cite{Ohoyama98} Neutrons with a wave length of ${\lambda}$$=$1.82646(6) ${\rm \AA}$ were obtained by the 331 Bragg reflection of a deformed singlecrystalline Ge monochromator, and 12$'$-blank-sample-27$'$ collimation. Sample powders were sealed in a $\phi$14.5mm cylindrical vanadium capsule in a standard aluminum cell with helium gas. The sample cell was mounted at the cold head of an Orange cryostat and cooled down to 1.3 K.

\begin{figure}[tb]
\begin{center}
\includegraphics[scale=0.4]{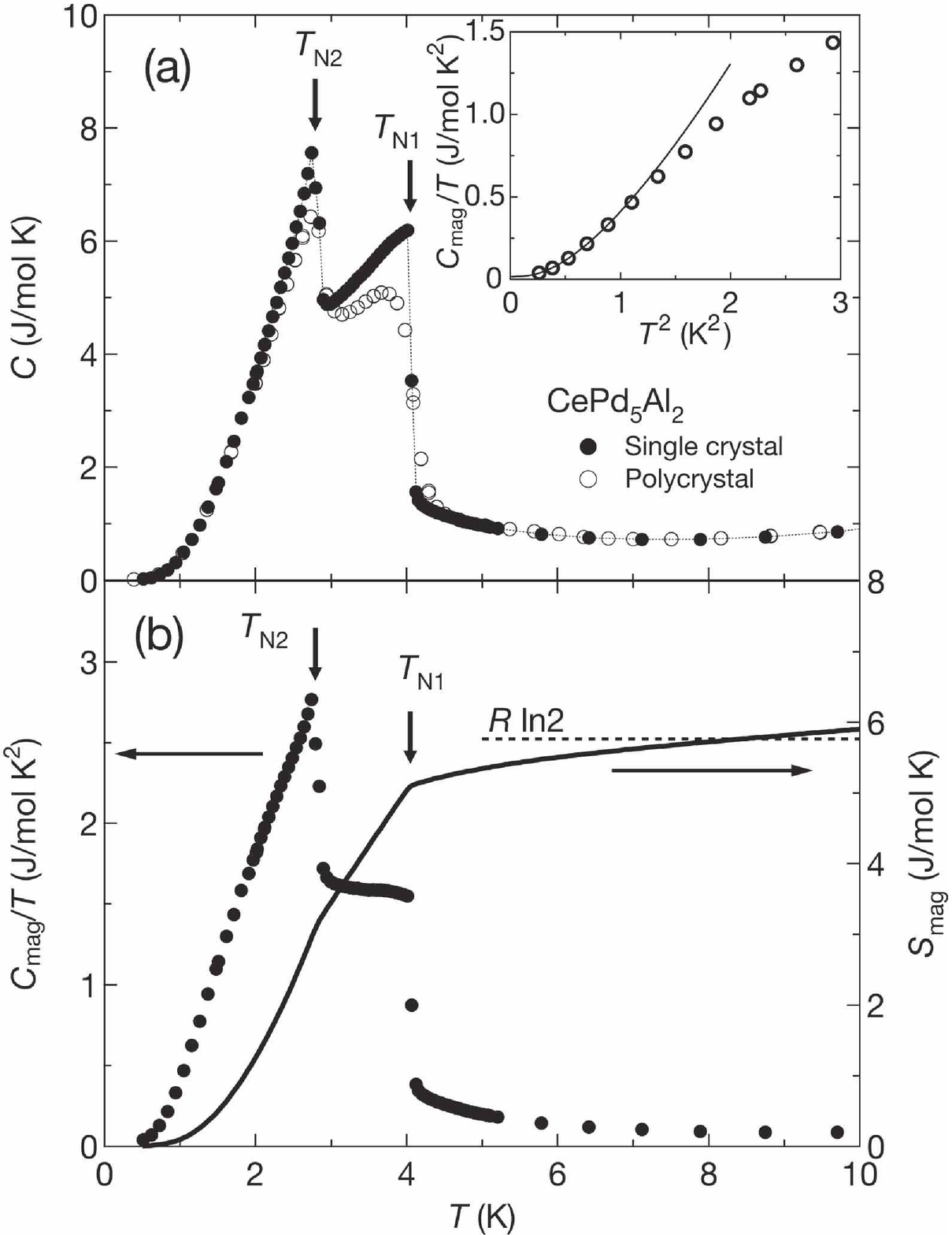}
\end{center}
\caption{(a) Temperature dependence of specific heat of the singlecrystalline {\cepdal} (œ) and the polycrystal (›).[16] Two anomalies appear at 4.1 K ({\tno}) and 2.9 K ({\tnt}). The inset shows $C_{\rm mag}$$/$$T$ vs $T^{2}$ plot. The electronic specific heat coefficient $\gamma$ is estimated as 18 mJ/mol K$^2$ by adopting the relation $C_{\rm mag}{=}{\gamma}{T}{+}{\beta}{T}^{3}e^{-{\Delta}_{\rm g}/k_{\rm B}T}$, which represents an antiferromagnteic magnon spectrum with a spin-wave energy gap ${\Delta}_{\rm g}$. (b) The temperature dependence of $C_{\rm mag}$$/$$T$ and the magnetic entropy $S_{\rm mag}$. At {\tno},  $S_{\rm mag}$ reaches 0.9$R$ln2.}
\label{f1}
\end{figure}

\begin{figure}[tb]
\begin{center}
\includegraphics[scale=0.4]{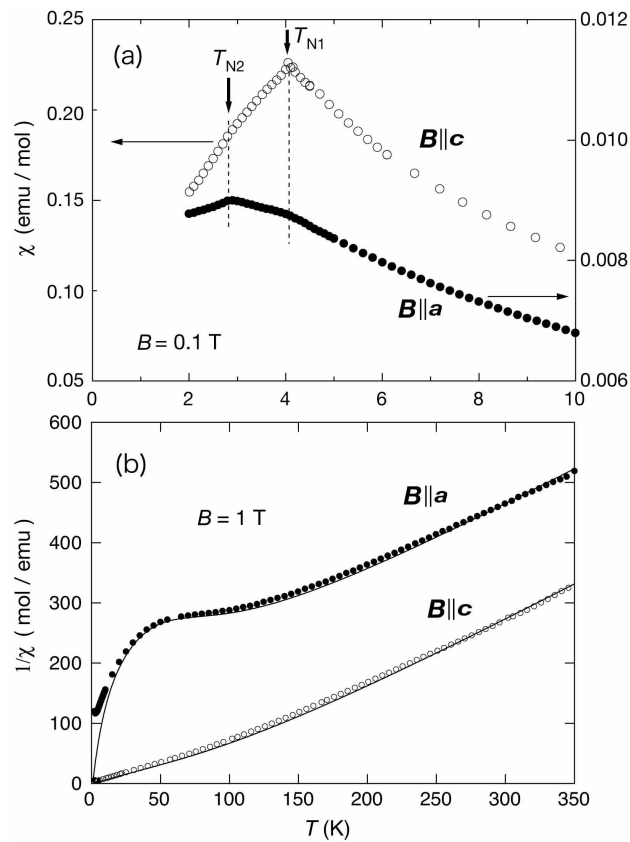}
\end{center}
\caption{(a) Temperature dependence of magnetic susceptibility ${\chi}$ in a magnetic field of $B$$=$0.1 T applied along the \textbf{\textit c}-  and  \textbf{\textit a}-axes. At 4 K, ${\chi}_{c}$ (left-hand scale) is 25 times as large as ${\chi}_{a}$ (right-hand scale), which means large uniaxial magnetic anisotropy in this system. ${\chi}_{c}$ shows a cusp at {\tno} and a broad shoulder at  {\tnt}, whereas ${\chi}_{a}$ shows a broad shoulder at  {\tno} and a cusp at {\tnt}.
(b) Temperature dependence of inverse magnetic susceptibility in a magnetic field of $B$$=$1 T applied along the \textbf{\textit c}- and \textbf{\textit a}-axes. Above 200 K, both ${\chi}_{c}^{-1}$ and ${\chi}_{a}^{-1}$ show the Curie-Weiss behavior. A significant difference in the paramagnetic Curie temperature ${\theta}_{p}$ and a broad shoulder in ${\chi}_{a}^{-1}$ around 50 K are due to the crystal electric field effect described in the text.}
\label{f2}
\end{figure}

\section{Results}
\subsection{Specific heat}

	Temperature dependence of the specific heat $C$($T$) of the singlecrystalline sample (closed circles) is compared with that of the polycrystal (open circles)[16] in Fig. \ref{f1}(a). The distinct peaks at 4.1 K and 2.9 K in the data of single crystal are much sharper than in the polycrystal. These temperatures are taken as {\tno} and {\tnt} because the peaks are the manifestation of successive magnetic transitions. The magnetic specific heat $C_{\rm mag}$ is evaluated by subtracting $C$($T$) of the polycrystalline {\ypdal} as the phonon contribution. By integrating $C_{\rm mag}$$/$$T$, the magnetic entropy $S_{\rm mag}$ was evaluated and plotted in Fig. \ref{f1}(b). At {\tno}, $S_{\rm mag}$ reaches 0.9$R$ln2, confirming that the CEF ground state should be a Kramers doublet. It also means that the Kramers doublet is not so strongly affected by the Kondo effect and magnetically orders at {\tno} and {\tnt}. Therefore, other degrees of freedom than the magnetic moments, for example, quadrupoles, are not involved in the transitions. Upon decreasing temperature below the AF transition temperatures, we expect the magnetic specific heat $C_{\rm mag}$ to follow the relation of $C_{\rm mag}{=}{\gamma}{T}{+}{\beta}{T}^3$, where ${\gamma}$ and ${\beta}$ are the coefficients  of an electronic and an AF magnon terms, respectively. However, as shown in the inset of Fig. \ref{f1}(a), there is no linearity in $C_{\rm mag}$$/$$T$ vs $T^2$. This discrepancy should arise from an AF magnon with a gap in the excitation spectrum due to anisotropy. As will be shown in the next subsection, in {\cepdal}, strong uniaxial anisotropy was found. Therefore, $C_{\rm mag}$ has an activation term, ie., $C_{\rm mag}{=}{\gamma}{T}{+}{\beta}{T}^{3}e^{-{\Delta}_{\rm g}/k_{\rm B}T}$, where ${\Delta}_{\rm g}$ is a spin-wave energy gap.\cite{Bredl87} Altogether, the fit to the data below 1 K is shown with the solid curve in the inset of Fig. \ref{f1}(a), which yields ${\gamma}{=}$18 mJ/mol K$^2$ and ${\Delta}_{\rm g}{=}$0.15 meV.  The rather low ${\gamma}$ value suggests the localized nature of the 4$f$ electrons in the ground state.

\subsection{Magnetic susceptibility and magnetization}

	Figure \ref{f2}(a) shows the temperature dependence of the magnetic susceptibility ${\chi}$ in a field of $B$$=$0.1 T applied along the \textbf{\textit c}{-} and \textbf{\textit a}{-}directionss. Note the different scale for ${\chi}_{c}$ (left-hand scale) and ${\chi}_{a}$ (right-hand scale). Strong uniaxial magnetic anisotropy is manifested in the large ratio ${\chi}_{c}$/${\chi}_{a}$$=$25 at {\tno}. Interestingly, ${\chi}_{c}$ shows a sharp cusp at {\tno} and a small knee at {\tnt}, whereas ${\chi}_{a}$ shows a small knee at {\tno} and a cusp at {\tnt}. Figure \ref{f2}(b) shows the temperature variation of ${\chi}_{c}^{-1}$ and ${\chi}_{a}^{-1}$ measured in a field of $B$$=$1 T. Above 200 K, both ${\chi}_{c}^{-1}$ and ${\chi}_{a}^{-1}$ are linear in $T$ as is expected from the Curie-Weiss law. From the slope, the effective magnetic moment is estimated as {\mp}$=$2.8 {\mb}$/$Ce-ion in both directions, whose value is close to 2.54 {\mb} expected for a trivalent Ce{-}ion. The paramagnetic Curie temperatures are ${\theta}_{p}{=}$37.8 and -159 K for \textbf{\textit B}$||$\textbf{\textit c} and \textbf{\textit B}$||$\textbf{\textit a}, respectively. As temperature decreases, ${\chi}_{a}^{-1}$ deviates from the Curie-Weiss law and passes through a broad shoulder at 50 K. The shoulder and the large difference in ${\theta}_{p}$'s should result from the CEF effect. We will present the results of the analysis using the CEF model afterwards.

\begin{figure}[tb]
\begin{center}
\includegraphics[scale=0.4]{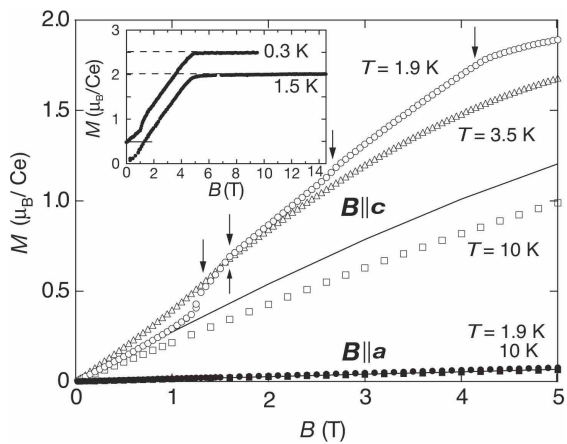}
\end{center}
\caption{Isothermal magnetization $M(B)$ in magnetic field $B$ applied along the \textbf{\textit c}- and \textbf{\textit a}-axes. Four anomalies at 1.4, 1.6, 2.8 and 4.2 T appear at 1.9 K ($T$$<${\tnt}), whereas one anomaly at 1.5 T appears at 3.5 K ({\tnt}$<$$T$$<${\tno}). No anomaly appears in $M$(\textbf{\textit B}$||$\textbf{\textit a}). The ratio $M$(\textbf{\textit B}$||$\textbf{\textit c})$/$$M$(\textbf{\textit B}$||$\textbf{\textit a}) is as large as 20, indicating the strong uniaxial magnetic anisotropy. The inset shows $M$(\textbf{\textit B}$||$\textbf{\textit c}) in magnetic field up to 14.5 T at 1.5 K (sample extraction method) and 9.5 T at 0.3 K (capacitance Faraday method). The data at 0.3 K is vertically shifted for clarity. The broken lines show the value of 2.04 {\mb}/Ce expected from the CEF scheme. }
\label{f3}
\end{figure}

	Figure \ref{f3} shows the magnetization isotherms ${M}{(}{B}{)}$ up to 5 T for \textbf{\textit B}$||$\textbf{\textit c} and \textbf{\textit B}$||$\textbf{\textit a}. At 1.9 K ($<${\tno}), $M_{c}$($B$) (open circles) exhibits four features at $B$$=$1.3, 1.5, 2.7 and 4.2 T, followed by a slow approach to saturation of 2.0 {\mb}/Ce at 14.5 T. (see in the inset of Fig. \ref{f3}) The saturation value is consistent with 2.04 {\mb}/Ce (broken line) which is expected from the CEF level scheme as is illustrated in Fig. \ref{f8}.  
At 3.5 K ({\tnt}${<}{T}{<}${\tno}), $M_c$ (open triangles) exhibits a small anomaly only at $B$$=$1.5 T. At 10 K ($>${\tno}), $M_c$ (open squires) increases monotonically with increasing magnetic fields. 
$M(B)$ at 0.3 K measured by the capacitance Faraday method is also shown in the inset of Fig. \ref{f3}. It starts increasing linearly with increasing magnetic field, followed by a rapid rise at $B_{\rm c1}{=}$1.0 T. Further, it exhibits an almost linear increase with increasing magnetic field above $B_{\rm c1}$ and reaches the saturation value of 2.0 {\mb}/Ce at $B_{\rm c2}{=}$4.9 T with a sharp knee. No apparent feature other than $B_{\rm c1}$ and $B_{\rm c2}$ could be detected because of variation of the magnetization data.  Apparently, $B_{\rm c1}$ and $B_{\rm c2}$ correspond to the critical fields of 1.3 and 4.2 T at 1.9 K, respectively, furthermore, the slope $M/B$ in the mediate range between $B_{\rm c1}$ and $B_{\rm c2}$ is temperature independent. Note that the extrapolation of $M(B)$ from the mediate range goes to ${M}{=}$0, suggesting that a magnetic structure above $B_{\rm c1}$ should be different from that below ${B}_{\rm c1}$.
By contrast, $M_a$ reaches only $\sim$0.1 {\mb}$/$Ce even at $B$$=$5 T and 1.9 K. The large ratio $M_c$$/$$M_a$$=$20 means that strong uniaxial magnetic anisotropy survives in the AF ordered state. 

\begin{figure}[tb]
\begin{center}
\includegraphics[scale=0.4]{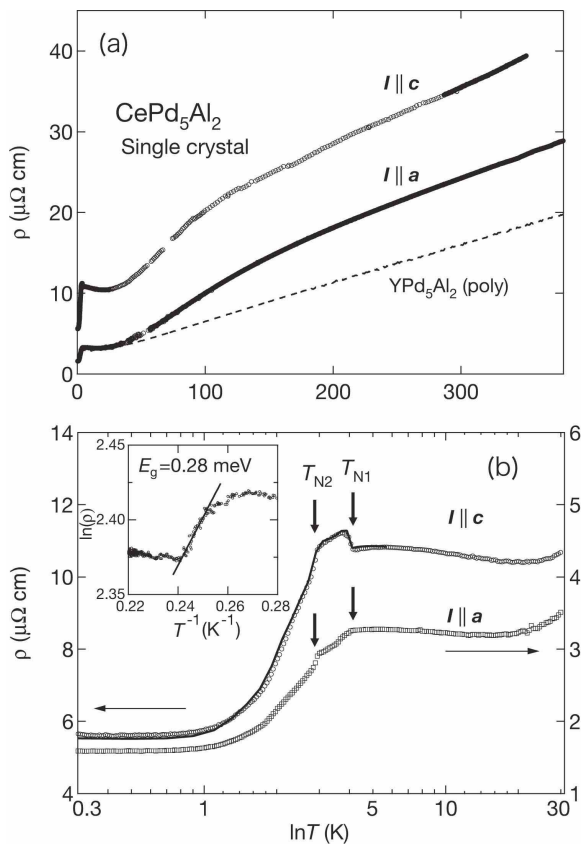}
\end{center}
\caption{(a) Temperature dependence of the electrical resistivity of a singlecrystalline {\cepdal} sample for current directions \textbf{\textit I}$||$\textbf{\textit c} and \textbf{\textit I}$||$\textbf{\textit a}. The broken line shows that of a polycrystalline {\ypdal} sample.
(b) Electrical resistivity ${\rho}$ for \textbf{\textit I}$||$\textbf{\textit c} (left-hand scale) and \textbf{\textit I}$||$\textbf{\textit a} (right-hand scale) vs ln$T$. The thick arrows indicate the transition temperatures. Between 6 K and 16 K, they obey $-$ln$T$ which results from the Kondo effect. The ${\rho}$($T$) for \textbf{\textit I}$||$\textbf{\textit c} jumps at {\tno} and suddenly drops at {\tnt} with decreasing temperatures. On the other hand, ${\rho}(T)$ for \textbf{\textit I}$||$\textbf{\textit a} bends at both {\tno} and {\tnt}. The bold solid curve is ${\rho}(T)$ for \textbf{\textit I}$||$\textbf{\textit c} estimated by Eq. (5) using the present data and some parameters. See text for details. The inset shows the plot of ln${\rho}$ vs 1${/}{T}$. The linear dependence gives the gap value of 0.28 meV for the superzone gap.}
\label{f4}
\end{figure}

\subsection{Electrical resistivity}

	Figures \ref{f4}(a) and \ref{f4}(b) show the temperature variations of the electrical resistivity ${\rho}$ of the singlecrystalline {\cepdal} for current directions \textbf{\textit I}$||$\textbf{\textit a} and \textbf{\textit I}$||$\textbf{\textit c} and that of the polycrystalline {\ypdal}. Both ${\rho}_{a}$($T$) and ${\rho}_{c}$($T$) decrease in parallel with decreasing temperatures down to 50 K. Thereby, the ration ${\rho}_{c}$/${\rho}_{a}$ increases from 1.4 at 300 K to 3.2 at 20 K. As is shown in Fig. \ref{f4}(b), ${\rho}_{a}$($T$) and ${\rho}_{c}$($T$) obey $-$ln$T$ dependence from 16 K to 6 K, which is a characteristic of the Kondo lattice compound. The slope is also anisotropic, $-$0.43 and $-$0.10 for \textbf{\textit I}$||$\textbf{\textit c} and \textbf{\textit I}$||$\textbf{\textit a}, respectively. On cooling below 5 K, ${\rho}_{c}$ jumps at {\tno}$=$4.1 K, peaks near 3.7 K, and bends at {\tnt}$=$2.9 K. The origin of the jump in ${\rho}_{c}$($T$) below {\tno} will be discussed later. On the other hand, ${\rho}_{a}$ shows knees at {\tno} and {\tnt} followed by a leveling at 1.6 ${\mu}{\Omega}$cm. The residual resistivity ratio (RRR) estimated from ${\rho}$(300 K)$/$${\rho}$(0.3 K) is 15 and 6 for \textbf{\textit I}$||$\textbf{\textit a} and \textbf{\textit I}$||$\textbf{\textit c}, respectively. 
	
\begin{figure}[tb]
\begin{center}
\includegraphics[scale=0.4]{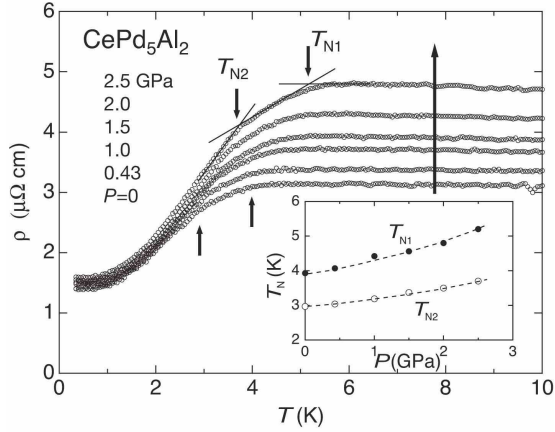}
\end{center}
\caption{Low temperature resistivity of a polycrystalline {\cepdal} sample under various pressures up to 2.5 GPa. The inset shows the pressure dependence of the ordering temperatures {\tno} and {\tnt}.}
\label{f5}
\end{figure}

	Pressure dependence of ${\rho}$(T) of polycrystalline {\cepdal} is represented in Fig. \ref{f5}. The drop at {\tno} and the kink at {\tnt} in the ${\rho}$(T) are indicated by the arrows at the cross-point of the two lines above and below the anomalies. With increasing pressure up to 2.5 GPa, both {\tno} and {\tnt} shift to higher temperatures as is shown in the inset of Fig. \ref{f5}. This fact suggests that {\cepdal} lies on the left-hand side of the gbell curveh in the Doniach phase diagram\cite{Doniach77} in which side the Kondo interaction is overwhelmed by the Ruderman-Kittel-Kasuya-Yosida (RKKY) interaction. It is certainly necessary to apply pressures above 3 GPa in order to search for the quantum critical behavior in this system.

\subsection{Neutron diffraction experiment}

\begin{figure}[tb]
\begin{center}
\includegraphics[scale=0.47]{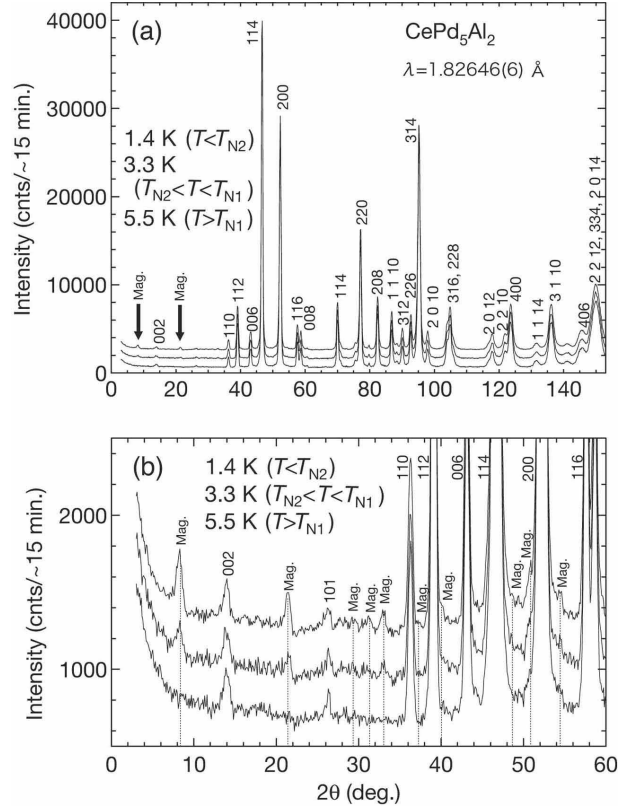}
\end{center}
\caption{(a) Powder neutron diffraction patterns at $T$$=$1.4 K ($T$$<${\tnt}), 3.3 K ({\tnt}$<$$T$$<${\tno}) and 5.5 K ($T$$>${\tno}). Each pattern is vertically offset for clarify. The thick arrows show magnetic reflections which appear below {\tno}. (b) The diffraction patterns expanded into the range 0${<}2{\theta}{<}$60$^{\circ}$ . The dotted lines show peak positions of magnetic reflections denoted as Mag. }
\label{f6}
\end{figure}

\begin{figure}[tb]
\begin{center}
\includegraphics[scale=0.4]{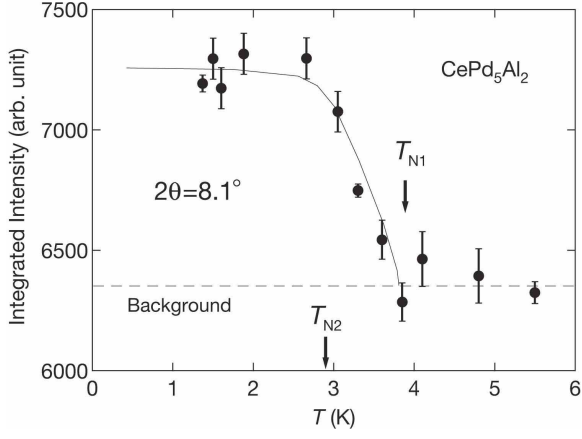}
\end{center}
\caption{Temperature dependence of integrated intensity of the magnetic peak at 2${\theta}{=}$8.1$^{\circ}$ corresponding to $|$\textbf{\textit Q}$|$$=$0.48  ${\rm \AA}^{-1}$. The solid line is a guide to the eyes. The broken line indicates the background intensity. The intensity of the magnetic reflection appears below {\tno}. It increases with decreasing temperatures and saturates at 2.6 K. No apparent change is found at around {\tnt}. }
\label{f7}
\end{figure}

	Figure \ref{f6}(a) shows the powder neutron diffraction patterns of {\cepdal} at the various temperatures 1.4, 3.3 and 5.5 K, corresponding to $T$$<${\tnt}, {\tnt}$<$$T$$<${\tno} and {\tno}$<$$T$, respectively. The patterns at 1.4 K and 3.3 K are vertically offset by 1000 and 2000 counts, respectively, for clarity. At 5.5 K, all the peaks can be indexed as nuclear Bragg peaks of the tetragonal ZrNi$_2$Al$_5$-type structure. No secondary phase was found in the patterns being consistent with the X-ray diffraction analysis. Although the intensities of the nuclear peaks do not change between 5.5 K and 1.4 K, weak superlattice Bragg peaks appear at 3.3 K and 1.4 K at the scattering angles  2$\theta$$=$8.1$^{\circ}$ and 21.7$^{\circ}$, which are shown with the thick arrows. The low angle patterns are expanded in Fig. \ref{f6}(b) to highlight the peaks labeled gMagh which manifest themselves at 3.3 K. Their intensities increase with decreasing temperature to 1.4 K while other Bragg peaks are unchanged. The integrated intensity of the superlattice Bragg peak at ${2}\theta$$=$8.1$^{\circ}$ corresponding to $|$\textbf{\textit Q}$|$=0.48 ${\rm \AA}^{-1}$ is plotted as a function of temperature in Fig. \ref{f7}. On cooling, the intensity starts to increase at 4 K, which is in good agreement with {\tno} determined by the macroscopic measurements. Thus the superlattice peaks originate in the magnetic transition at {\tno}. Since no apparent anomaly exists at {\tnt}, a propagation vector of magnetic structures should not change so much at {\tnt}, which hints to a possible magnetic structure. From the temperature dependence of the magnetic susceptibility shown in Fig. \ref{f2}(a), one possibility is canting of the magnetic moments from the \textbf{\textit c}-direction. The other is square-up of a modulated sine-wave structure with appearance of higher order reflections, because the change of the magnetic entropy at {\tnt} is very large as is seen in the specific heat. 
However, no sign of the higher order reflections was observed in the present experiments. This is probably because the intensity was too weak to detect by means of the powder neutron diffraction technique.

\section{Discussion}

\subsection{CEF analysis}
\begin{figure}[tb]
\begin{center}
\includegraphics[width=5.5cm]{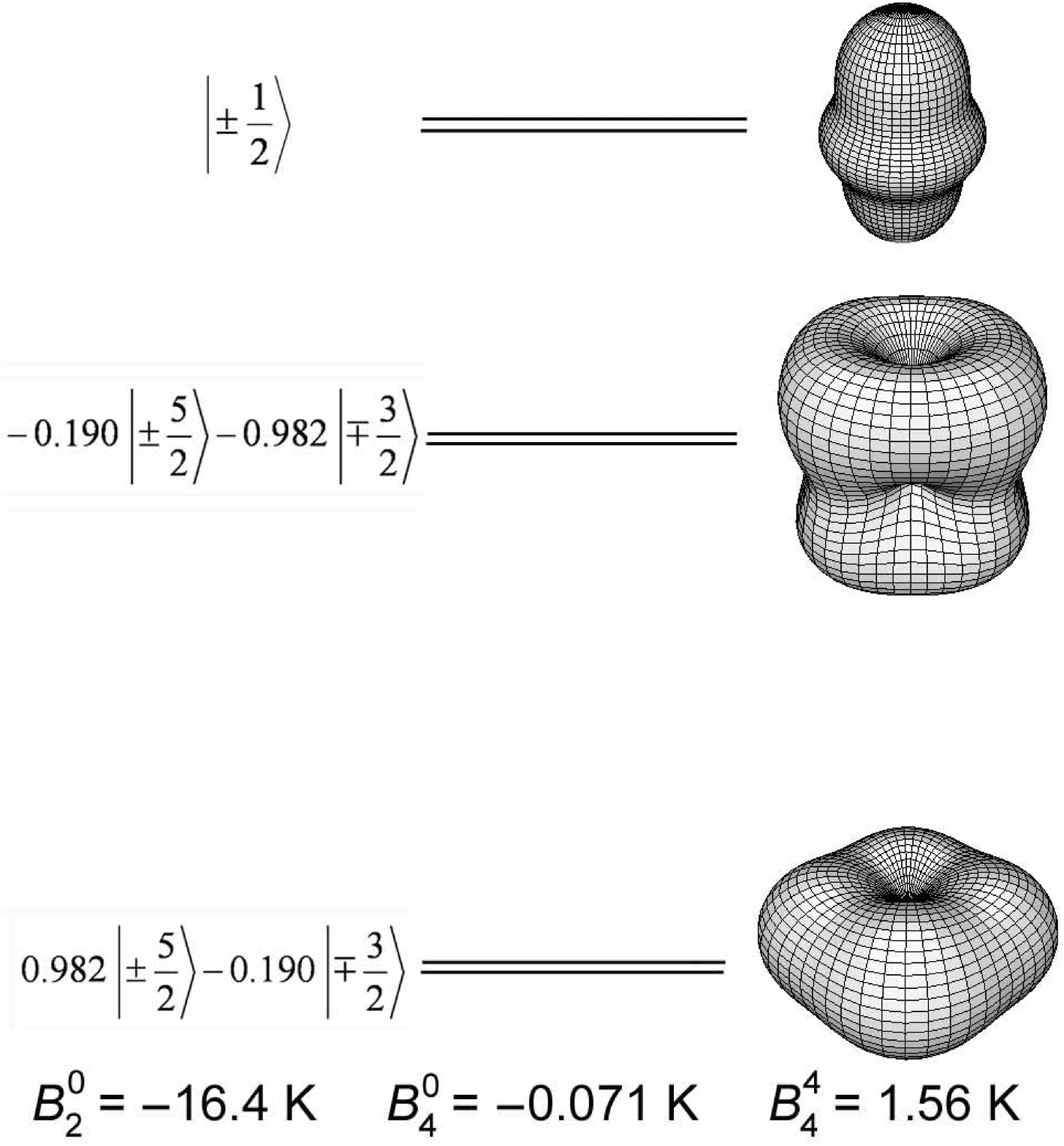}
\end{center}
\caption{CEF level scheme and the wave functions, the spatial distributions of the wave functions of the 4$f$ electron in {\cepdal} calculated by using CEF parameters. The wave function of the CEF ground state consists primarily of $\left | {\pm}\frac{5}{2} \right \rangle$ and the excitation levels are far from the ground state level ($\Delta_{1}$$=$230 K and $\Delta_{2}$$=$300 K), indicating that the magnetic moments should be forced to align along the \textbf{\textit c}-axis.}
\label{f8}
\end{figure}

	We first analyze the data of ${\chi}(T)$ and $M$($B$) in the paramagnetic state by using the CEF model. The local symmetry surrounding the Ce ions in {\cepdal} is represented by the crystallographic point group of $D_{4h}$, hence, the CEF Hamiltonian can be written as
\begin{equation}
H_{\rm CEF}=B_{2}^{0} O_{2}^{0}+ B_{4}^{0} O_{4}^{0}+ B_{4}^{4} O_{4}^{4}	
\end{equation}
where $B_m^n$ and $O_m^n$ stand for the CEF parameters and the Stevens operators, respectively.\cite{Hutching64} The above Hamiltonian splits a multiplet of Ce$^{3+}$ with ${J}{=}\frac{5}{2}$ into three Kramers doublets. In the mean-field approximation, temperature dependence of magnetic susceptibility ${\chi}_i$($T$) (${i}{=}a, c$) is expressed as
\begin{equation}
	{\chi}_i(T) = {\chi}_{i{\rm CEF}}(T) / [1-{\lambda}{\chi}_{i{\rm CEF}}(T)] 
\end{equation}	
where $\lambda$ is the mean-field parameter representing the exchange interaction among Ce magnetic moments. The CEF contribution to the magnetic susceptibility, ${\chi}_{i{\rm CEF}}$($T$), is given by
\begin{multline}
{\chi}_{i \rm CEF} = \frac{(g_{J} {\mu}_{\rm B})^{2}}{\displaystyle\sum\limits_{m} e^{-E_{m}/k_{\rm B}T}} 
 \Biggl( \frac{\displaystyle\sum_{m}{\langle}m{|}J_{i}{|}m{\rangle^{2} e^{-E_{m}/k_{\rm B}T}}}{k_{\rm B}{T}} \\
+ \sum_{m} \sum_{n({\neq}m)} {\langle}n{|}J_{i}{|}m{\rangle}^{2}  \frac{e^{-E_{m}/k_{\rm B}T} - e^{-E_{n}/k_{\rm B}T}}{E_{n} - E_{m}}  \Biggr)
\end{multline}	
where the Land${\rm \acute{e}}$ $g$ factor $g_J$ is 6/7 for Ce$^{3+}$, $J_i$ is the $i$-component of the angular momentum, $m$ and $n$ are eigenstates of 4$f$ wave functions, and $E_{m(n)}$ is the eigenvalue of the derived CEF level.  The CEF contribution to the magnetization, $M_{\rm iCEF}$, is given by
\begin{equation}
M_{i{\rm CEF}} = g_{J} {\mu}_{\rm B} \sum_{m} {\langle}m{|}J_{i}{|}m{\rangle}\frac{e^{-E_{m}/k_{\rm B}T}}{\displaystyle\sum\limits_{m} e^{-E_{m}/k_{\rm B}T}}
\end{equation}	 
A least-square fit to the data ${\chi}(T)$ was done by setting the second order CEF parameter $B_2^0$ to be $-$20.5 K as an initial value, which was estimated from the difference between $\theta_{p}$(\textbf{\textit I}$||$\textbf{\textit c}) and $\theta_{p}$(\textbf{\textit I}$||$\textbf{\textit a}). The determined CEF parameters $B_m^n$ and $\lambda$ are listed in Table I. By using these parameters, ${\chi}_{a}(T)$ and ${\chi}_{c}(T)$ are calculated and drawn by solid curves in Fig. \ref{f2}(b). It should be noted that they agree well with the data in the paramagnetic state. Furthermore, the calculated curves $M$($B$) are drawn by solid curves in Fig. \ref{f3}. At 10 K, the larger values than the experimental data may result from either AF correlation between the Ce ions even in the paramagnetic state or suppression of the magnetic moments due to the Kondo effect. Figure \ref{f8} shows the CEF energy level scheme and the wave functions. The first and second excited levels lie at 230 and 300 K above the ground state whose wave function consists primarily of $\left | {\pm}\frac{5}{2} \right \rangle$ as shown in Table \ref{t1}. The large uniaxial magnetic anisotropy in the paramagnetic state should originate from the isolated CEF ground state. Moreover, the pancake-like spatial distribution of the wave functions supports the magnetic moments to orient along the \textbf{\textit c}-axis.

\begin{table}[tb]
\caption{Crystalline electric field parameters, molecular field coefficient, energy levels and wave functions of the 4$f$ electron in {\cepdal} determined from the analysis of the magnetic susceptibility ${\chi}_{a}(T)$ and ${\chi}_{c}(T)$, and magnetization $M$(\textbf{\textit B}$||$\textbf{\textit a}) and $M$(\textbf{\textit B}$||$\textbf{\textit c}) in the paramagnetic state.}
\label{t1}
\begin{tabular}{ccccccc}
\hline
CEF&$B_{2}^{0}$ (K) & $B_{4}^{0}$ & $B_{4}^{4}$ & \multicolumn{2}{c}{$\lambda$ (mol/emu)} & \\
parameters& -16.4 & -0.071 & 1.56  & \multicolumn{2}{c}{-2.13} & \\
\hline
\hline
Energy& \multicolumn{6}{l}{} \\
levels& \multicolumn{6}{l}{Wave functions} \\
\hline 
\vspace{-2mm}\\
$E$ (K) & $\left | {+}\frac{5}{2} \right \rangle$ &$\left | {+}\frac{3}{2} \right \rangle$ &$\left | {+}\frac{1}{2} \right \rangle$ &$\left | {-}\frac{1}{2} \right \rangle$   & $\left | {-}\frac{3}{2} \right \rangle$ & $\left | {-}\frac{5}{2} \right \rangle$ \\
\vspace{-2mm}\\
\hline
300 & 0 & 0 & 1 & 0 & 0 & 0 \\
300 & 0 & 0 & 0 & 1 & 0 & 0 \\
230 &0.190& 0 & 0 & 0 &0.982& 0 \\
230 & 0 &0.982& 0 & 0 & 0 &0.190\\
0 &0.982& 0 & 0 & 0 &-0.190& 0 \\
0 & 0 &-0.190& 0 & 0 & 0 &0.982\\
\hline
\end{tabular}
\end{table}

\subsection{Superzone gap formation}

We now focus on the anisotropic behavior of ${\rho}(T)$ near {\tno} and {\tnt}. The fact that ${\rho}_{c}(T)$ jumps below {\tno} but ${\rho}_{a}(T)$ shows a small kink may be explained by two scenarios: (i) a spin density wave (SDW) transition and (ii) AF superzone formation. The SDW transition occurs to reduce the electron energy by nesting flat Fermi surfaces. Examples are Cr\cite{Fawcett88,McWhan67}, Ce(Ru$_{0.85}$Rh$_{0.15}$)$_2$Si$_2$\cite{Miyako96,Murayama97}, Ce$_7$Ni$_3$\cite{Umeo96}, URu$_2$Si$_2$\cite{Broholm87}, YbPtBi\cite{Movshovich94}, etc, where the 3$d$ or 4$f$(5$f$) electrons  have itinerant character to be involved in the Fermi surfaces. However, the 4$f$ electrons in {\cepdal} are well localized as indicated from the small $\gamma$ value of 18 mJ/mol K$^2$. The localize nature is also supported by the weak effect of pressure on {\tno} and {\tnt}. Therefore, the scenario of a SDW transition is discarded. 
	
	A superzone gap can open in a certain direction on the Fermi surface when localized 4$f$ electrons order antiferromagnetically. Because ${\rho}_{a}(T)$ of {\cepdal} shows no enhancement at {\tno}, the superzone gap should open along the \textbf{\textit c}-axis. This is reasonable because the length of the reciprocal vector ${|}\textbf{\textit c}^{\ast}{|}$ is much smaller than that of ${|}\textbf{\textit a}^{\ast}{|}$, providing higher possibility of opening a superzone gap along the \textbf{\textit c}-direction. The configuration is similar to what was observed in Er metal\cite{Green61,Mackintosh62}, where AF and ferromagnetic transitions occur at $T_{1}{=}$85 and $T_{2}{=}$19.6 K, respectively. Below $T_{1}$, enhancement of ${\rho}_{c}(T)$ is a consequence of the AF ordering with a magnetic structure characterized by longitudinal modulation moments. The propagation vector was determined to be \textbf{\textit k}$\sim$[0, 0, $\frac{2}{7}$]\cite{Gibbs86}, resulting in formation of new zone boundaries.
	
	Following the procedure to evaluate the gap energy in Er, we estimate the energy gap in {\cepdal} from the thermal activation-type form of ${\rho}$$=$${\rho}{'}e^{E_{\rm g}/k_{\rm B}T}$, where ${\rho}{'}$ is an initial resistivity, $E_{\rm g}$ is the energy gap and $k_{\rm B}$ is the Boltzmann$'$s constant. The solid line in the inset of Fig. \ref{f4}(b) yields $E_{\rm g}$ as 0.28 meV (3.2 K). Furthermore, we calculated ${\rho}_{c}(T)$ and plotted the result with the bold solid curve in Fig. \ref{f4}(b). The total resistivity can be expressed as follows,\cite{Miwa62,Elliott63,Elliott64}
\begin{equation}
{\rho}_{\rm total} = \frac{\rho_{0} + \rho_{\rm ph} + \rho_{\rm mag}}{1 - gm(T)}
\end{equation}	  
where ${\rho}_0$ is a residual resistivity, and ${\rho}_{\rm ph}$ and ${\rho}_{\rm mag}$ are phonon and magnetic contributions to the resistivity, respectively. We evaluated ${\rho}_{\rm ph}$ from the $T$-dependent part of the electrical resistivity of the polycrystalline {\ypdal}, and assumed ${\rho}_{\rm mag}$ to be scaled by 3.8 times as large as the $T$-dependent part of ${\rho}_{a}$. In Eq. (5), $m(T)$ is the temperature dependence of sublattice magnetization in the ordered state. In the present calculation, $m(T)$ is assumed to be $m(T){=}\{1{-}({T}{/}{T}_{\rm N})^{2}\}^{1/2}$  which is the same as the BCS-gap. Although the translation factor $g$, which scales the effect of $m(T)$, depends on the band gap and shape of the Fermi surface, it is left as a parameter here. Nevertheless, this simple model with ${\rho}_0$$=$ 4.72 ${\mu}{\Omega}$ cm and $g$$=$0.18 can reproduce well the complex behavior of ${\rho}_{c}(T)$ as shown in Fig \ref{f4}(b). On further cooling, the appearance of the hump in ${\rho}_{c}(T)$ at {\tnt} is a result of competition between the enhancement due to the superzone gap and the reduction due to the development of the magnetic ordering. 

\subsection{Magnetic structures and magnetic anisotropy in the AFM states}

	We now focus our attention on the magnetic structures and the strong anisotropy in the AF ordered state. The strong anisotropy of ${\chi}(T)$ and $M(B)$ such as ${\chi}_{c}$$/$${\chi}_{a}$$=$25 in 0.1 T and at 4 K, $M_{c}$$/$$M_{a}$$=$20 in 5 T and at 1.9 K, respectively, means that the magnetic moments are forced to orient along the \textbf{\textit c}-direction. From a perspective on the strong uniaxial anisotropy in the magnetic property, it is instructive to compare the magnetic properties of {\cepdal} with those of a dense Kondo compound CeNi$_2$Al$_5$.\cite{Isikawa91,Isikawa94} This compound crystallizing in a body centered orthorhombic PrNi$_2$Al$_5$-type structure undergoes an AF transition at $T_{\rm N}$$=$2.6 K. It shows large uniaxial anisotropy with the easy direction along the \textbf{\textit b}-direction; ${\chi}_{b}$$/$${\chi}_{c}$$=$27 and ${\chi}_{b}$$/$${\chi}_{a}$$=$8 at 2.6 K, $M_b$$/$$M_c$$=$27 and $M_b$$/$$M_a$$=$9 in 5 T and at 1.4 K, respectively. The CEF ground state is a Kramers doublet, and the first and second excitation levels lie at 210 and 650 K above the ground state, respectively. In the AF ordered state, the magnetization isotherm $M(B)$ exhibits anomalies only for \textbf{\textit B}$||$\textbf{\textit b} in 3.9 and 6.1 T at 1.4 K, whereas those for \textbf{\textit B}$||$\textbf{\textit a} and \textbf{\textit B}$||$\textbf{\textit c} increase monotonically with increasing magnetic field. 
Neutron diffraction experiments reveal that the magnetic structure can be characterized by the propagation vector  \textbf{\textit k}$=$[0.500,0.405,0.083], where the magnetic moments align with a sine-wave modulated structure. 
The anisotropy and the jump at $B_{\rm a}$ in the magnetization curves in the AF ordered state could not be explained by the simple two-sublattice mean-field model, however, the periodic mean-field model\cite{Blanco91,Gignoux93} with isotropic magnetic interaction $J({Q}{=}|\textbf{\textit k}|)$ between the Ce-ions and its higher-order terms could reproduce well the anisotropic magnetization isotherm including the anomalies in $M$(\textbf{\textit B}$||$\textbf{\textit b}).\cite{Givord96} In the case of {\cepdal}, the magnetic structure should be a modulated one characterized by the propagation vector with $|\textbf{\textit k}|$$=$0.48 ${\rm \AA}^{-1}$. Therefore, the periodic mean-field model could explain the multi-step magnetization isotherm $M$(\textbf{\textit B}$||$\textbf{\textit b}) when the magnetic structure could be determined. We are now in progress to perform neutron diffraction experiments on a single crystal. 
	
	To the best of our knowledge, the uniaxial anisotropy in the magnetic properties of {\cepdal} is the strongest among the tetragonal Ce intermetallic compounds. The anisotropy is much larger than that of the well studied orthorhombic system CeZn$_2$, which was suggested to be a model system of the 3D-Ising model.\cite{Gignoux92} Various ground states can emerge in the 3D-Ising system due to competing magnetic interactions. Moreover, they are strongly influenced by magnetic field, and thus show multistep metamagnetic transitions, complex magnetic phase diagrams and a tricritical point in the diagram. Such phenomena were observed in some AF systems like CeZn$_2$\cite{Gignoux92}, FeBr$_2$\cite{Katsumata97}, Co(SCN)$_2$(CH$_3$OH)$_2$\cite{DeFotis07}, etc. Because {\cepdal} exhibits multistep metamagnetic transitions and a complicated magnetic phase diagram, it is a candidate for the 3D-Ising systems in the 4$f$ electron system.

\section{Conclusions}

We performed electrical resistivity ${\rho}$, magnetic susceptibility ${\chi}$, magnetization $M$ and specific heat measurements on the singlecrystalline {\cepdal}. The successive AF transitions at {\tno}$=$4.1 K and {\tnt}$=$2.9 K were confirmed. The magnetic entropy reaches a value close to $R$ln2 at {\tno}, indicating that the relevant Kramers state is not so much affected by the Kondo effect. The localized nature of the 4$f$ electron state was indicated by the small $\gamma$ value of 18 mJ/mol K$^2$. The magnetization measurements revealed strong uniaxial anisotropy; ${\chi}_{c}$$/$${\chi}_{a}$$=$25 in 0.1 T and at 4 K, $M_c$$/$$M_a$$=$20 in 5 T and at 1.9 K. The giant uniaxial anisotropy observed in both the paramagnetic state and AF ordered state should result from the CEF ground state. The analysis of the CEF model revealed that the 4$f$ electron ground state has a wave function consisting primarily of the $\left | {\pm}\frac{5}{2} \right \rangle$ components and its energy level is isolated from the first and second excited state by 230 and 300 K. The magnetization isotherm exhibits several anomalies only in \textbf{\textit B}$||$\textbf{\textit c}; 1.3, 1.5, 2.7 and 4.2 T at 1.9 K ($<${\tno}), and 1.5, 2.7 T at 3.5 K ({\tnt}$<$$T$$<${\tno}), whereas no anomaly emerges in \textbf{\textit B}$||$\textbf{\textit a} up to 5 T. These properties are very similar to those of CeNi$_2$Al$_5$, the orthorhombic system with comparably strong uniaxial anisotropy. The anisotropic properties including the magnetization isotherm for CeNi$_2$Al$_5$ were well understood using the periodic mean-field model with the isotropic magnetic interactions between Ce moments. The close similarity indicates that the strong uniaxial anisotropy in {\cepdal} in the AF ordered state should arise from the CEF effect in the presence of isotropic interactions. 

	The electrical resistivity also shows anisotropic behavior such as ${\rho}_{c}$$/$${\rho}_{a}$$=$3.2 at 20 K. At {\tno}=4.1 K, the substantial jump only in ${\rho}_{c}$ suggests the formation of a superzone gap along the \textbf{\textit c}-direction. The resistivity measurements were done on the polycrystal sample under pressures up to 2.5 GPa. Thereby, both {\tno} and {\tnt} were increased slowly, indicating that this system should be dominated by the RKKY interaction. To access a quantum critical point, measurements under pressures above 3 GPa are needed. Powder neutron diffraction experiments revealed the development of magnetic peaks at $|$\textbf{\textit Q}$|$$=$0.48 ${\rm \AA}^{-1}$ below {\tno}. This observation indicates a long-periodic modulated structure but does not allow determination of the propagation vector. Though no additional change appears below {\tnt}, the magnetic structure should change at {\tnt} because apparent anomalies were found there in the macroscopic measurements. A small change of the propagation vector or evolution of an anti-phase structure may occur with appearance of the high-order reflections.

\section*{Acknowledgments}

The authors would like to thank F. Iga for helpful discussions.
They also thank Y. Shibata for the electron-probe microanalysis performed at N-BARD, Hiroshima University and K. Nemoto in IMR, Tohoku University, for helpful assistance in the neutron diffraction experiments. 
The electrical resistivity, magnetization and specific heat measurements were carried out at N-BARD, Hiroshima University. This work was financially supported by a Grant-in-Aid for Scientific Research (A), No. 18204032, from the Ministry of Education, Culture, Sports, Science and Technology.

\end{document}